\documentclass[num-refs]{nbdt-article}

\usepackage{siunitx}

\usepackage{amsmath, amsfonts}
\usepackage{amssymb}
\usepackage{booktabs}
\usepackage{multirow}
\usepackage{float}
\usepackage{hyperref}
\usepackage{graphicx}
\usepackage{authblk}

\papertype{Original Article}
\paperfield{Journal Section}

\title{The Influence of Width Ratios on Structural Beauty in Male Faces}

\author[1]{Theresa Tennstedt}
\author[1]{Benjamin Knopp}
\author[1]{Dominik Endres}

\affil[1]{Department of Psychology, Marburg University, Germany}

\corraddress{Theresa Tennstedt, Department of Psychology, Marburg University, Germany}
\corremail{theresa.tennstedt@uni-marburg.de}

\runningauthor{Tennstedt et al.}

\begin{document}

\maketitle

\begin{abstract}
This study investigates the relationship between interocular distance relative to overall facial width (width ratio) and perceived subjective beauty in male faces. Building on the methodology of \cite{Pallett2010}, who found that average proportions in female faces were rated as most attractive, the current study aimed to test this hypothesis in male faces. Faces from the Chicago Face Database \citep{Ma2015} were morphed into average faces within three groups (with low, medium, and high width ratios), each composed of 96 or 97 individual images. These three average faces were then systematically manipulated in their width ratios across three levels in both directions, respectively, resulting in a total of 21 comparable faces. The use of multiple base faces served as a control for potential artifacts of image processing. Consequently, comparisons were restricted to within-group pairs to avoid confounding by co-varying facial features (e.g., skin tone), which precluded direct cross-condition comparisons but ensured internal validity. In a two-alternative forced-choice task, participants selected the more beautiful face from each pair. The data were analyzed using a Bayesian model which enables inference of the width ratio perceived as most beautiful. Results support the hypothesis that averageness in facial proportions correlates with higher perceived attractiveness. The study highlights the importance of controlling for image manipulation, including attempts at methodological implementation, and of considering ethnicity as a potential moderating variable. These findings offer a data-driven foundation for understanding facial aesthetics and cognitive processes of human perception, with applications in advertising, artificial face generation, and plastic surgery.

% Please include a maximum of seven keywords
\end{abstract}

%Hypothesen:
%1. Welche Gesichter werden bei Männern am schönsten empfunden? -> UV: Breitenverhältnis, AV: empfundene Schönheit
%2. Sind die schönsten Verhältniss bei Männern die gleichen wie bei Frauen? -> deskriptiv oder auf Populationsebene untersuchen?
%3. Sind auch bei Männern die schönsten Verhältnisse die durchschnittlichen? -> deskriptiv oder auf Populationsebene untersuchen?

%Main text
\section{Introduction}
\label{Introduction}

% \subsection{The Perception of Beauty}
% \label{The perception of beauty}

The predominance of subjective preferences is generally taken for granted when it comes to the perception of beauty. The extent to which this is truly the case, or whether beauty can also be assessed objectively, is a question that humanity has been pondering for centuries \citep{Singer2024, Dietrich2022, Vashi2015, Edler2001}. The golden ratio, which dates back to antiquity and was widely applied by Renaissance artists such as da Vinci and Dürer \citep{Edler2001}, represents one such attempt to identify a universal pattern in natural structures that may evoke a sense of beauty and that has since been repeatedly reproduced in art and architecture \citep{Singer2024, Harrar2018}. It arises when the ratio of a total length to its longer segment is the same as the ratio of the longer segment to the shorter one, yielding a value of approximately \mbox{$\phi = \frac{1+\sqrt 5}{2} \approx 1.618$}. Golden ratios have not only been observed in nature and reproduced in objects (buildings, paintings, symbols, etc.), but have historically represented the ideal human body proportions \citep{Breiner2019} and have been applied to facial proportions as well \citep{Ricketts1982}.

Today, however, the assumption that the golden ratio forms the basis of perceived beauty is considered outdated \citep[e.g.,][]{Naini2024}. Contemporary research on the predictability of facial beauty at the population level has identified additional influencing factors, including symmetry, sexually dimorphic facial features, averageness, and similarity to one's own face \citep{Langlois1990, Little2011, Vashi2015, Sarwer2003, Borelli2010}. As with other topics in psychological perception, beauty has become a cross-disciplinary field of study, investigated on behavioral, cognitive, and neurobiological levels \citep[e.g.,][]{Skov2021, Raggio2022, Dietrich2022, Yang2022}.

\subsection{The Social Dimensions of Beauty}
\label{The Social Dimensions of Beauty}

Beyond biological and cognitive frameworks, beauty should also be understood as a social concept that is shaped by cultural norms, historical context, and media influences \citep{Adams1977, Raggio2022}.
Cross-cultural studies demonstrate significant variation in beauty ideals, challenging the notion of universal standards \citep[e.g.,][]{Zhang2019, Sorokowski2013}. Furthermore, perceived attractiveness functions as a form of social capital, influencing a wide range of interpersonal judgments and outcomes, including perceived trustworthiness, competence evaluations, and hiring decisions \citep{Jackson1995, OConnor2018, Zebrowitz2008, Frevert2014, Hosoda2003, Mobius2006}. This phenomenon can be explained by the halo effect, whereby positive aesthetic impressions generalize to other attributed traits \citep{Thorndike1920, Dion1972, Langlois2000}. Consequently, judgments of beauty are not made in a vacuum but are embedded in, and reinforced by, societal structures and learned preferences \citep{Adams1977, Eagly1991}.

\subsection{Preference for Average Faces}
\label{Preference for average faces}

To investigate a potential universal component within this socially complex landscape, this study focuses on one of the most replicable effects in facial attractiveness research: the preference for average faces \citep{Dimitrov2023, Rhodes1999}. This preference allows for the operationalization of beauty through measurable geometric properties, such as facial width or height ratios. Preferences for average faces can be explained by two different theoretical approaches. According to the evolutionary biological perspective, individuals with average faces are perceived as healthier and more developmentally stable \citep[e.g.,][]{Pallett2010, Rhodes2006}. Because attractiveness is thought to serve a reproductive function, faces associated with these qualities are preferred over those that deviate from the average. At the same time, evolutionarily influenced social preferences that are shaped by individual experiences and learned expectations, introduce variability in facial preferences \citep{Little2011}.

According to the cognitive theory, individuals form an internal prototypical face, representing the average of all the faces they have encountered. Newly perceived faces are compared to this prototype, and those that resemble it are preferred because they require less cognitive effort to process, making their internal representation more efficient in terms of mental energy expenditure \citep{Langlois1990, Pallett2010}. This cognitive preference can be explained by perceptual fluency, that is, the ease of processing prototypical stimuli, which in turn elicits a positive response \citep{Alter2009, Langlois1990, Arkes1991}. The inherent familiarity of average faces further enhances the overall processing fluency, creating a self-reinforcing cycle that solidifies the preference for facial averageness \citep{Juravle2024, Little2011}.

In their study \textit{The Influence of Each Facial Feature on How we Perceive and Interpret Human Faces} \citet{Diego-Mas2020} provided further evidence that faces are processed holistically rather than as a collection of individual components, especially in comparison to other objects, \citep[cf. also][]{Tanaka1993}. They argue that in this process, local facial features play a subordinate, context-dependent role and are only one part of the overall, global perception of the face. However, differences seem to appear in the perception of dynamic versus static faces \citep{Rubenstein2005}. Other studies have shown that facial processing and the perception of facial attractiveness in particular are influenced by gender-specific factors. This effect is evident with respect both to the gender of the observer and the gender of the face being evaluated \citep[e.g.,][]{Qi2022, Zhang2016}.

In the field of psychophysics, modern research seeks to examine the influence of the physical facial proportions on perception \citep{Fan2012}. Considering both evolutionary biological and cognitive approaches, the question arises as to what extent perceived facial attractiveness is related to their geometric features. In their study "New 'Golden' Ratios For Facial Beauty", \cite{Pallett2010} investigated the relationship between physical facial proportions, averageness, and perceived attractiveness. To do this, they systematically varied the width and length ratios of female faces. The width ratio was defined as the distance between the pupils relative to the total width between the inner edges of the ears, while the length ratio was defined as the distance between the eyes and the mouth relative to the total face length from hairline to chin. Through assessment of the attractiveness of the different faces, they found that the optimum ratios of 46\% for width and 36\% for length corresponded closely to the average proportions found in women's faces, which \cite{Pallett2010} referred to as the 'new golden ratios'.

    %\citeauthor{Danel2007} deliver a study on sexual dimosphism in face features, assuming the eye-mouth-eye angle being a gender-specific feature. They could find an association between smaller angles and the trait of masculinity in male faces which makes them being perceived more attractive by women \citep{Danel2007}. To what extent the length and width ratios found in women by \citep{Pallett2010} are applicable to male faces should be investigated in this study.

\subsection{Aim of This Study}
\label{Aim of this study}

To prevent biases and distortions resulting from the assumed gender specificity of facial perception and attractiveness, it is important to conduct studies on facial perception that include participants of all genders and examine both male and female faces.
%Hypotheses
Building upon the findings of \cite{Pallett2010}, that were limited to the investigation of female faces, the present study aims to test the hypothesis that the preference for average ratios can also be observed in male faces. We assume that, similar to female faces, male faces with average proportions are perceived as most beautiful. However, given the preference for gender-specific traits, we furthermore hypothesize that the optimal ratios in male faces might differ from those identified in female faces. More precisely, this study focuses on the variation of width ratios while keeping other facial features constant.

%Methods
\section{Methods}
\label{Methods}

To investigate the relationship between geometric facial proportions and perceived beauty, the methodology of \cite{Pallett2010} was adapted. The aim was to determine which proportions are perceived as particularly beautiful in male faces and whether these correspond to the proportions identified for female faces \citep{Pallett2010}. For this purpose, male faces from a dataset of 290 images were morphed to create average composite faces, whose width ratios were systematically manipulated.

The present study focused exclusively on the investigation of width ratios, as a full replication of \cite{Pallett2010} was not feasible due to insufficient methodological detail regarding the manipulation of length ratios. The width ratio was defined as the distance between the pupils relative to the total face width, measured as the distance between the inner edges of the ears. The resulting face stimuli were presented in a two-alternative forced choice task to $n = 97$ participants. Participants were asked to compare two simultaneously displayed faces with different width ratios and to indicate via keypress which face they intuitively perceived as more beautiful. Data analysis was conducted using Bayesian statistical methods.
%ggf. hier schon Methodik spezifizieren

\subsection{Generation of Faces and Manipulation of Their Width Ratios}
\label{Generation of faces}

\begin{figure}[!htbp]
\centering
\includegraphics[width=0.8\textwidth]{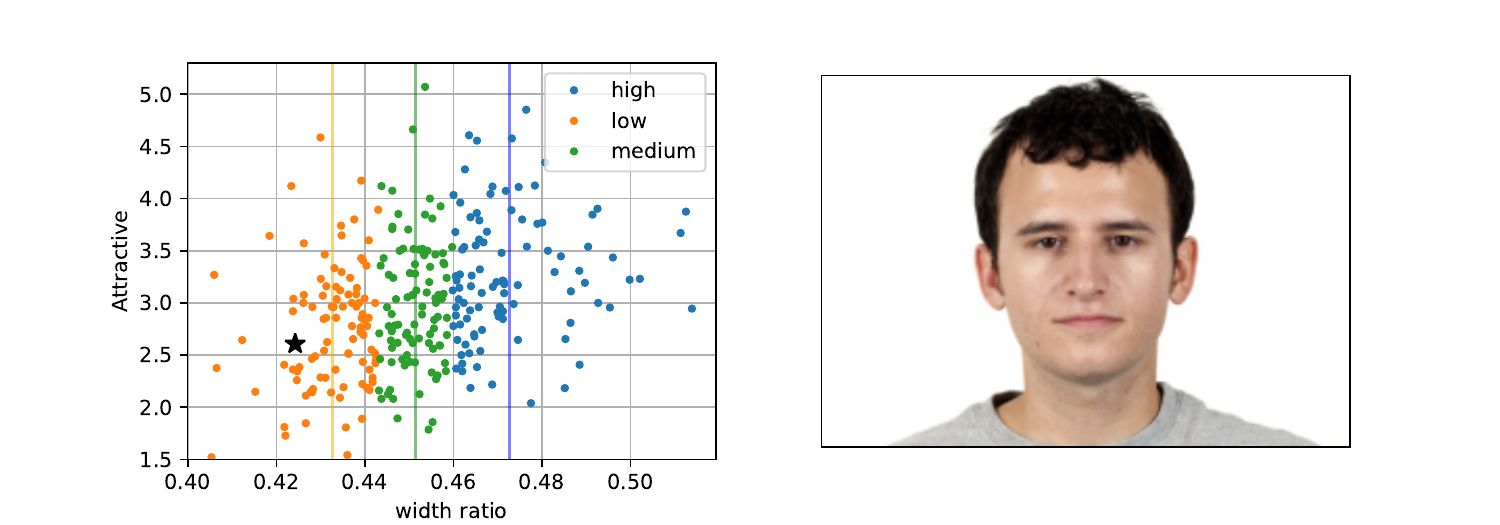}
\caption{\textbf{Left:} Scatterplot of faces from the Chicago Face Database, with attractiveness ratings on the y-axis and width ratios on the x-axis. Colors indicate group membership: \textit{low} (orange), \textit{medium} (green), and \textit{high} (blue) width ratios, defined by dividing the dataset into thirds based on width ratio values. Vertical lines represent the mean width ratio of each group. Note: The attractiveness ratings provided by the database were not used in this study’s assessment of beauty. \textbf{Right:} Example image from the Chicago Face Database (marked with a star in the plot).}
\label{fig: plot_database}
\end{figure}

\begin{figure}[!htbp]
\begin{minipage}[h]{\textwidth}
\centering
\includegraphics[width=0.4\textwidth]{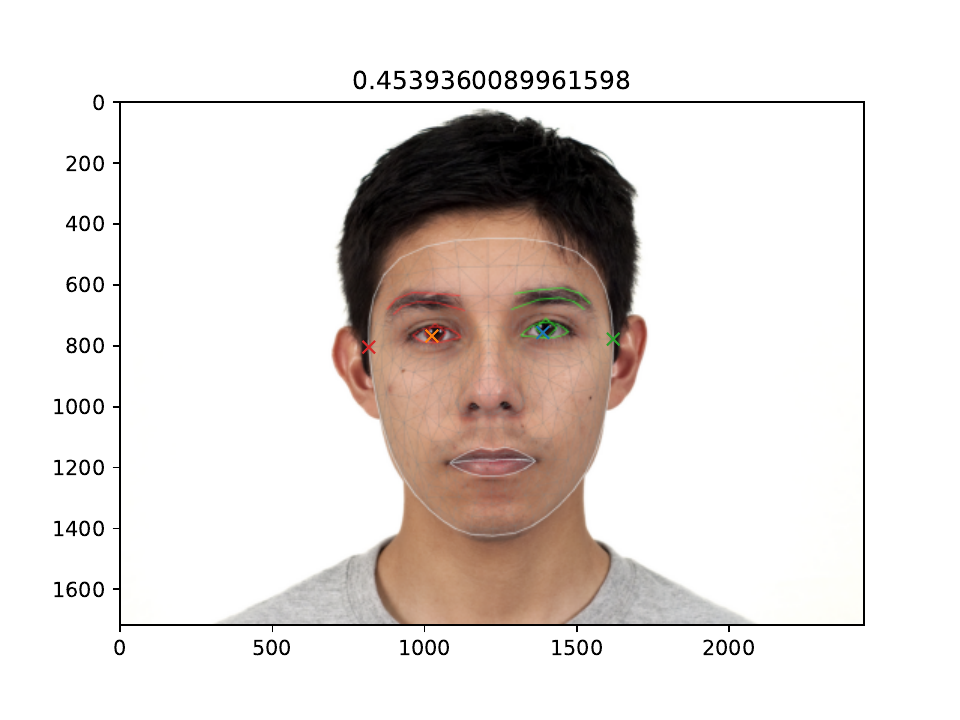}
\includegraphics[width=0.4\textwidth]{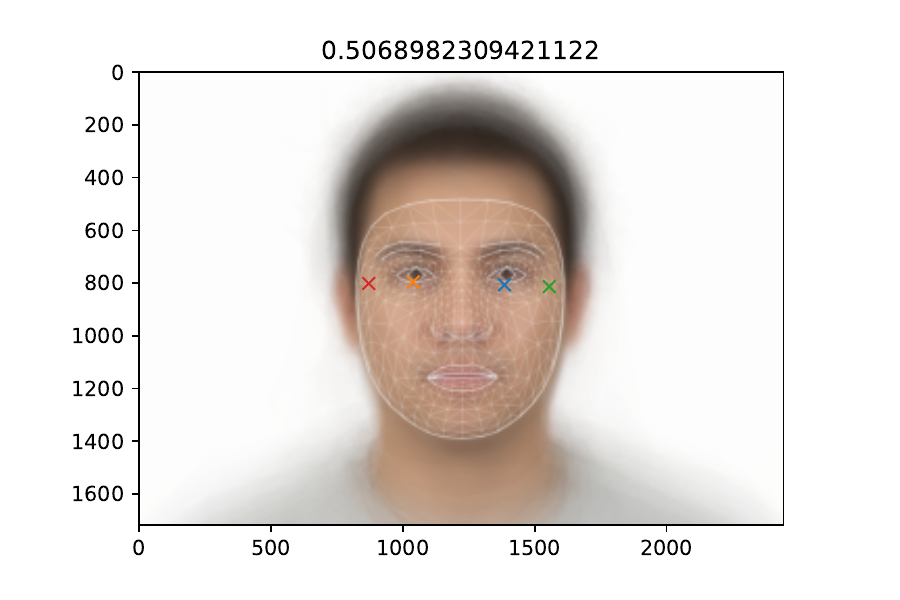}
\end{minipage}
\caption{\textbf{Left:} Facial template showing the measurement points used to calculate the width ratio, defined as the ratio of inner eye distance to total face width. The width ratio of the depicted face is indicated above the image. \textbf{Right:} Morphing mask used to generate the average faces, shown here overlaid on the composite face of the \textit{low} width-ratio group.}
\label{fig: facemask}
\end{figure}

\begin{figure}[!htbp]
\begin{minipage}[h]{\textwidth}
\centering
\includegraphics[width=0.3\textwidth]{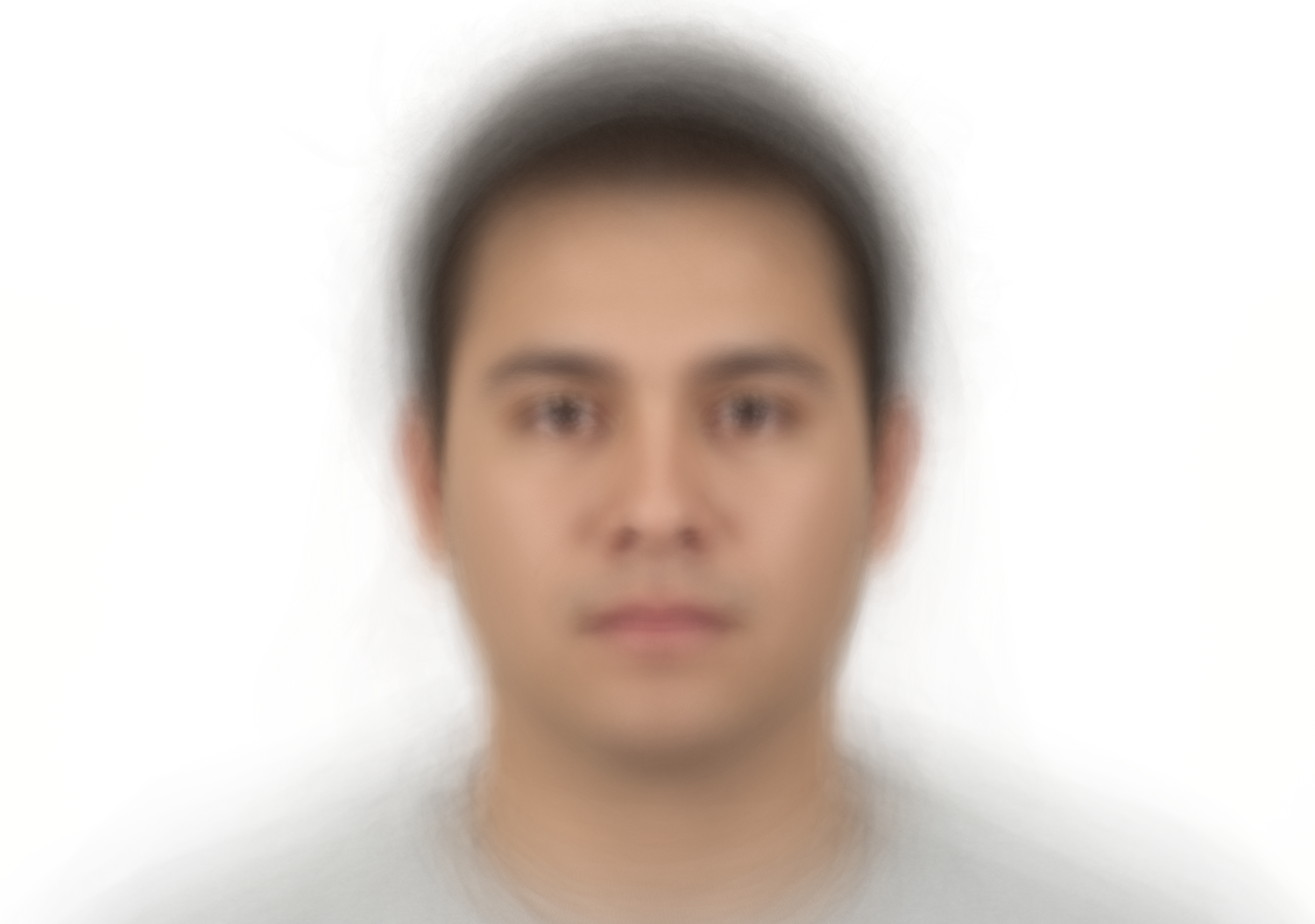}
\includegraphics[width=0.3\textwidth]{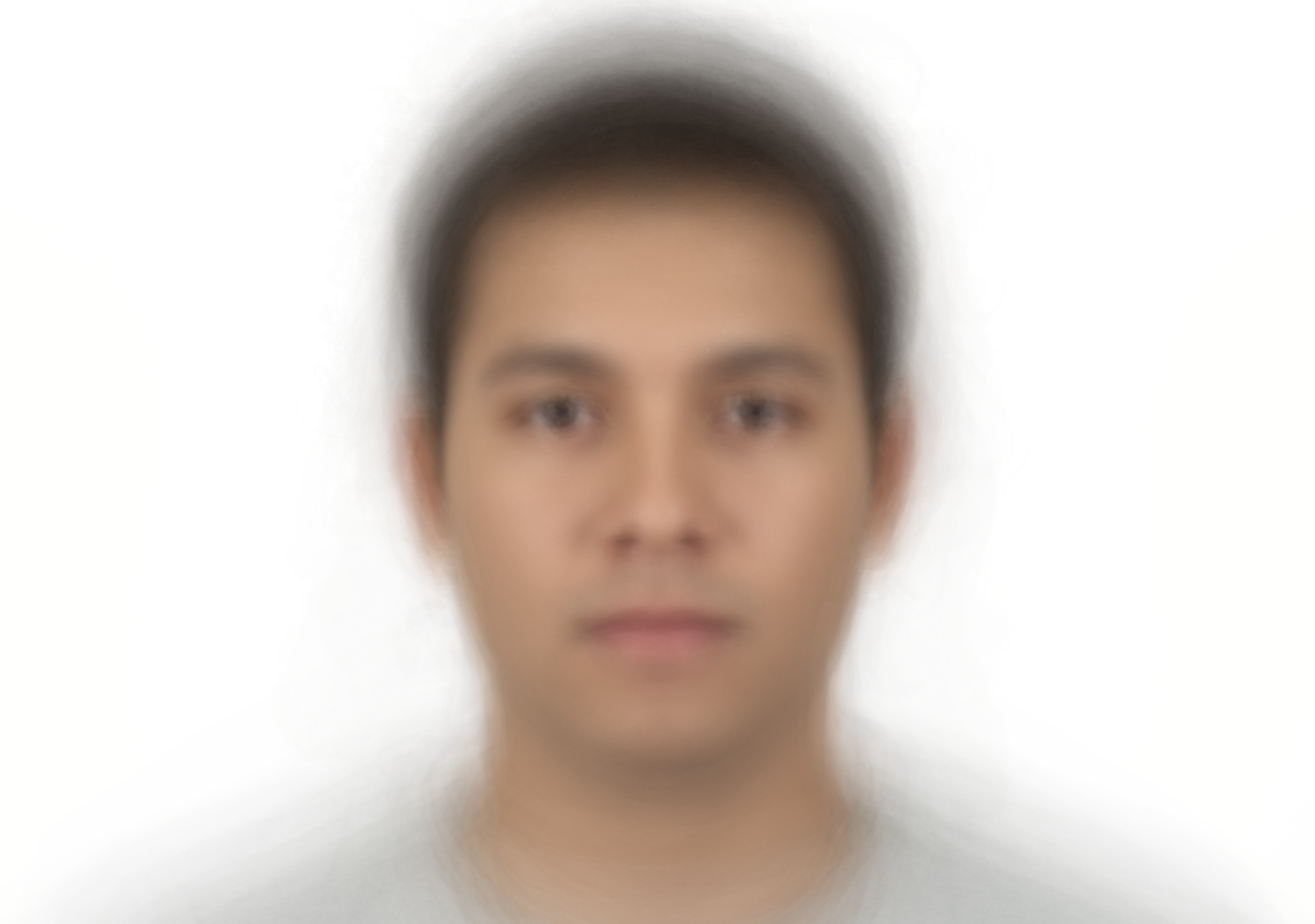}
\includegraphics[width=0.3\textwidth]{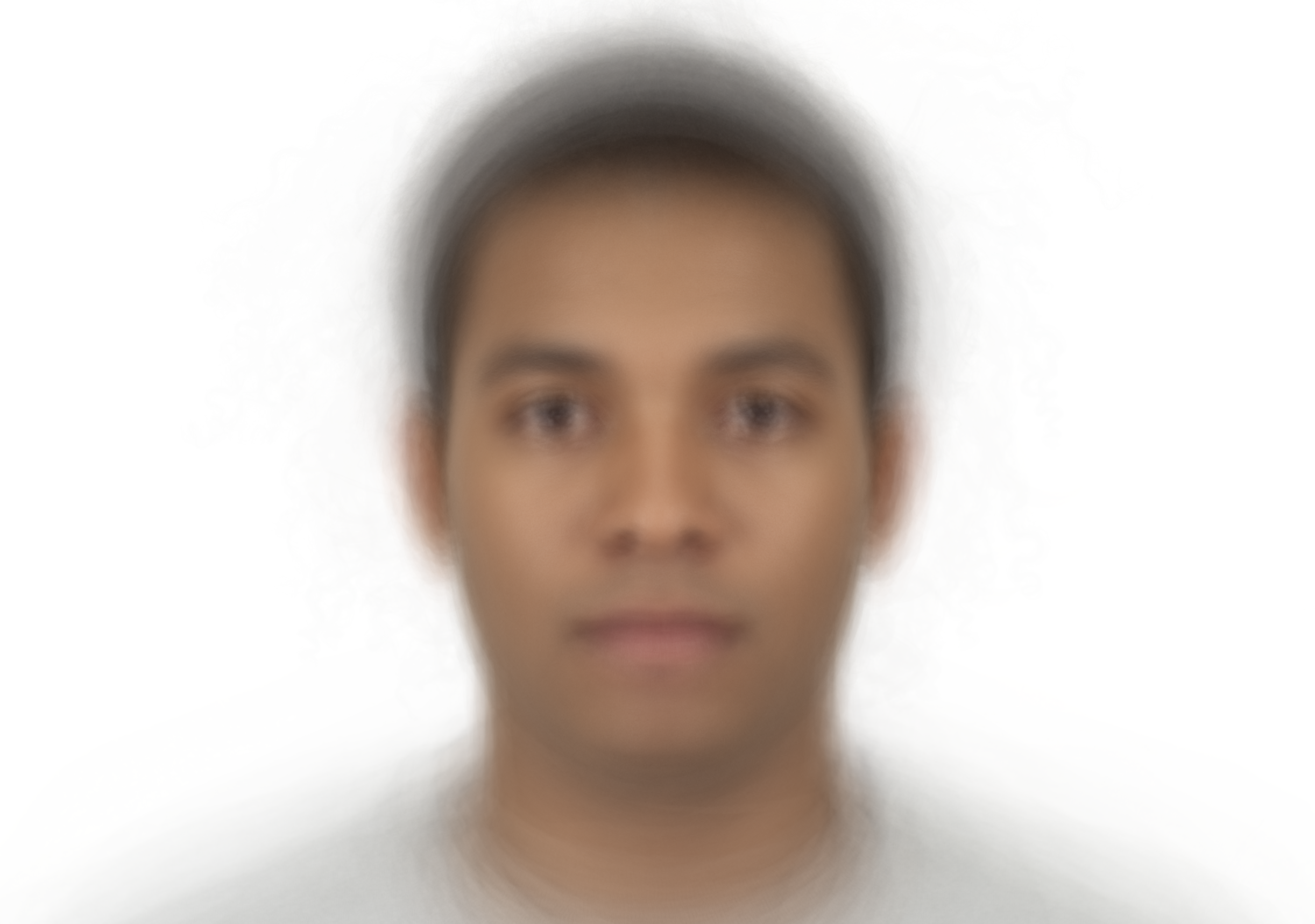}
\end{minipage}
\caption{Average composite faces created from one-third of the male faces in the Chicago Face Database grouped by ratio: \textit{low} (left), \textit{medium} (middle), and \textit{high} (right).}
\label{fig: average faces}
\end{figure}

To manipulate the width ratio of the male faces and ensure comparability across stimuli, all 290 images from the Chicago Face Database \citep{Ma2015} were grouped into three equally sized categories depending on their original width ratios (Figure \ref{fig: plot_database} ). Images from each group were then morphed to create composite faces by computing the mean of the RGB channels of every pixel (Figure \ref{fig: facemask} ).
This morphing process eliminated individual differences in facial features, resulting in natural-looking stimuli with consistent facial features across all trials. As a result, three average faces were generated, each composed of 97, 96, and 97 individual faces, respectively. These correspond to the experimental conditions labeled \textit{low}, \textit{medium}, and \textit{high}, based on their average width ratios (Figure \ref{fig: average faces} ).

%Figure: plot

%Figure: facemasks

%Figure: average faces

\begin{figure}[!htbp]
\begin{minipage}[h]{\textwidth}

(a)
\includegraphics[width=0.11\textwidth]{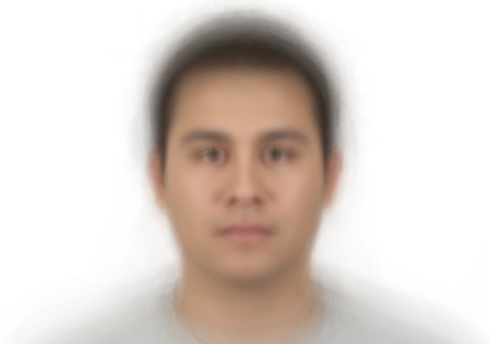}
\includegraphics[width=0.11\textwidth]{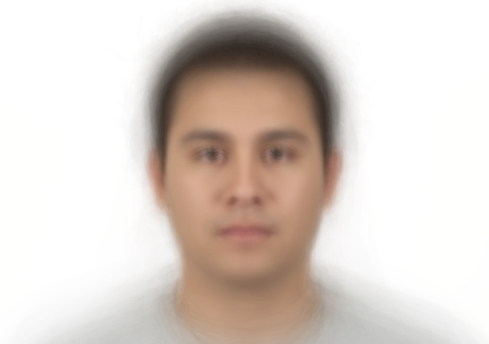}
\includegraphics[width=0.11\textwidth]{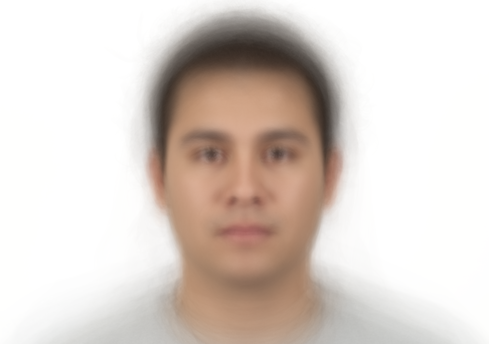}
\includegraphics[width=0.11\textwidth]{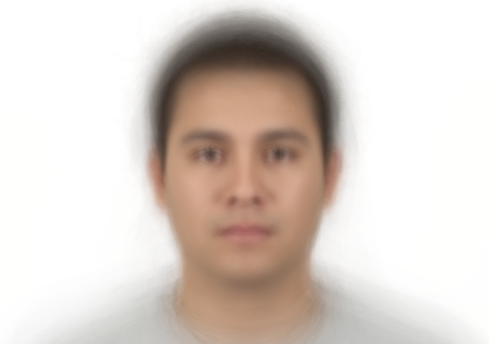}
\includegraphics[width=0.11\textwidth]{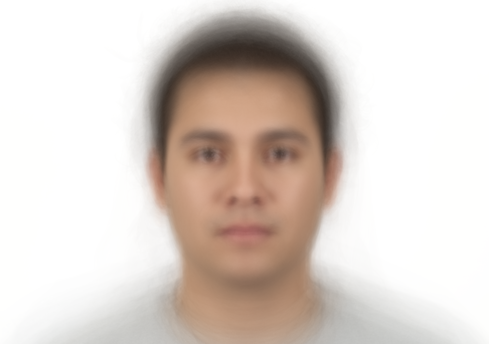}
\includegraphics[width=0.11\textwidth]{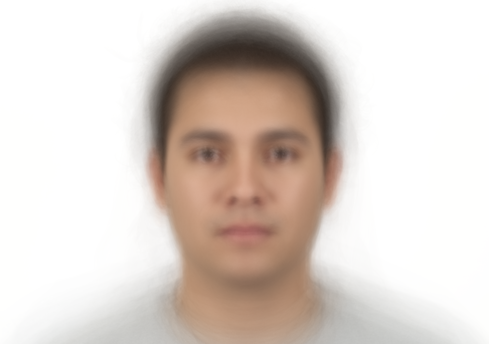}
\includegraphics[width=0.11\textwidth]{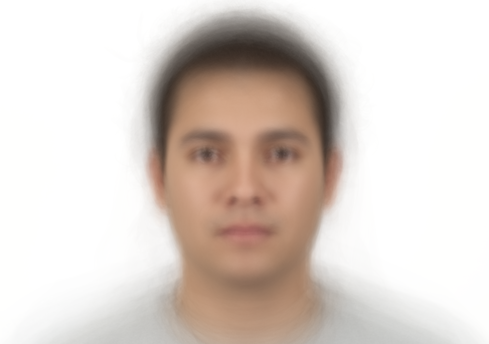}
(b)
\includegraphics[width=0.11\textwidth]{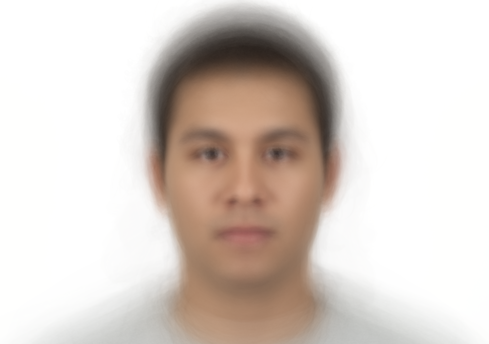}
\includegraphics[width=0.11\textwidth]{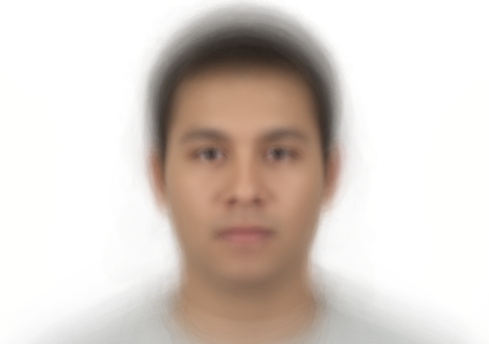}
\includegraphics[width=0.11\textwidth]{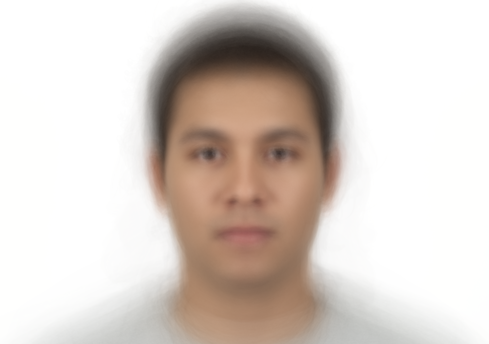}
\includegraphics[width=0.11\textwidth]{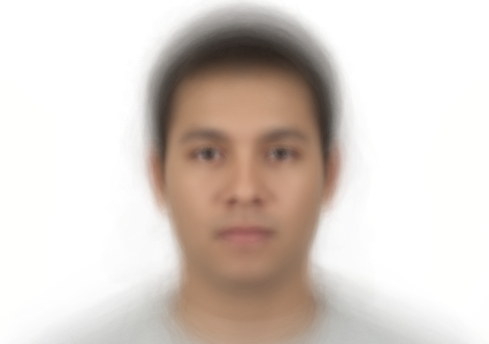}
\includegraphics[width=0.11\textwidth]{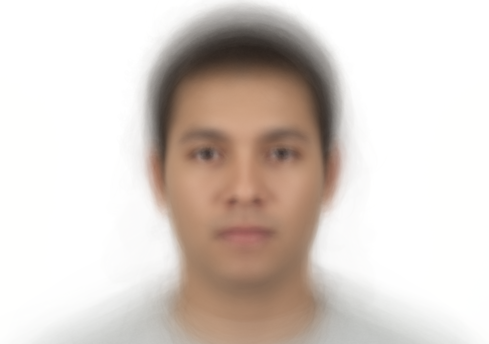}
\includegraphics[width=0.11\textwidth]{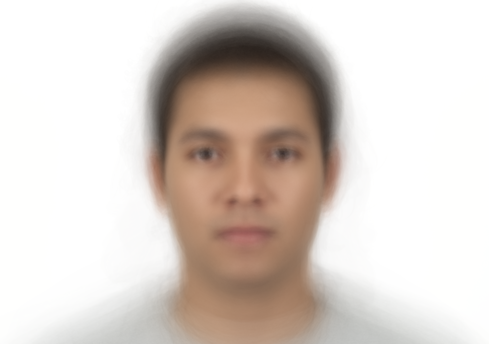}
\includegraphics[width=0.11\textwidth]{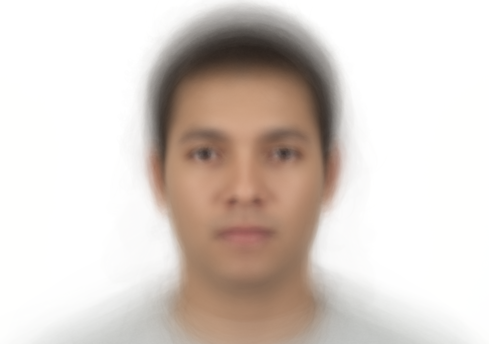}
(c)
\includegraphics[width=0.11\textwidth]{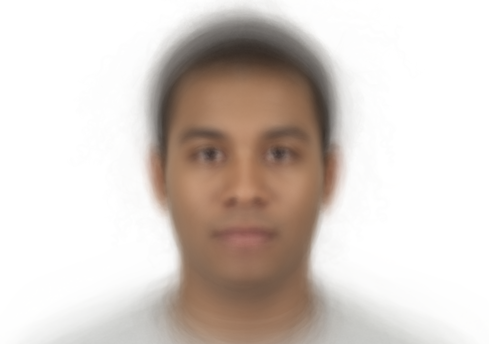}
\includegraphics[width=0.11\textwidth]{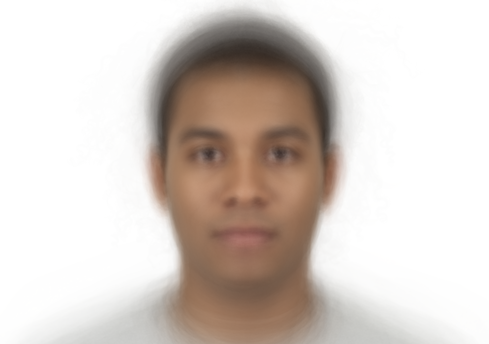}
\includegraphics[width=0.11\textwidth]{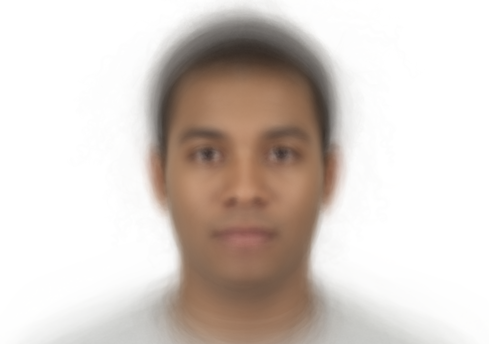}
\includegraphics[width=0.11\textwidth]{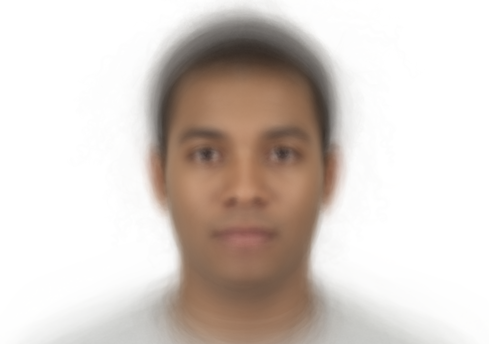}
\includegraphics[width=0.11\textwidth]{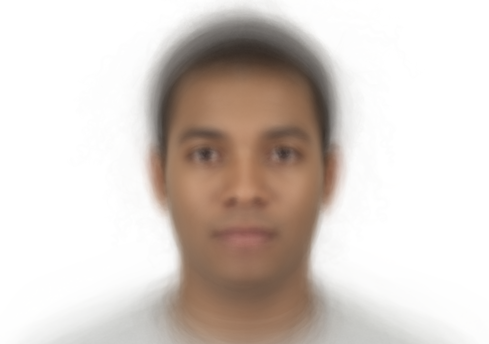}
\includegraphics[width=0.11\textwidth]{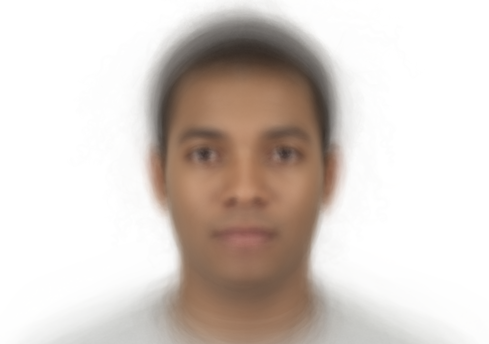}
\includegraphics[width=0.11\textwidth]{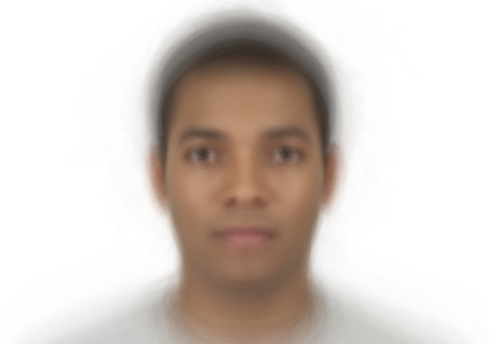}
\includegraphics[width=0.5\textwidth]{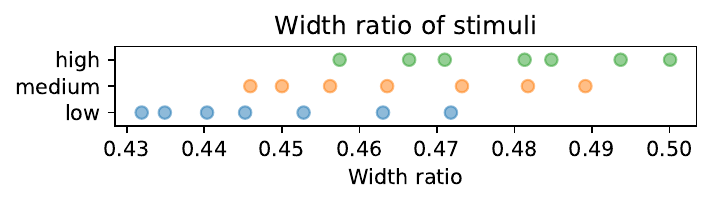}
\end{minipage}
    \caption{Overview of all stimuli used in the experiment, consisting of 21 manipulated faces generated by morphing 97, 96, and 97 images from the \textit{low} (a), \textit{medium} (b), and \textit{high} (c) width ratio groups, respectively. From left to right in each condition: $-100$, $-66$, $-33$, $0$, $+33$, $+66$, $+100$ (arbitrary units). The original average face ($0$) served as the baseline and was manipulated to generate the other faces. (d) The diagram below shows the program-calculated width ratios of all faces, verifying the consistency and validity of the manipulations.}
\label{fig: manipulated faces}
\end{figure}

The resulting images of the three different average faces were subsequently manipulated in their width ratios using Adobe Photoshop \cite{Inc.2023}. The Liquefy filter was applied to systematically adjust the eye distance in six incremental steps: -100, -66, -33, +33, +66, and +100 arbitrary units (provided by the Liquefy filter). By doing so, the distance between the eyes was either decreased or increased three times at equal intervals, respectively, resulting in six manipulated faces and one unaltered original for each condition. An overview of all manipulated stimuli and their calculated width ratios is shown in Figure \ref{fig: manipulated faces}.

%figure: manipulated faces

The purpose of grouping the faces into the three conditions (\textit{low}, \textit{medium}, and \textit{high} width ratios) was to control for potential effects of image editing on the evaluation of beauty. It was hypothesized that the same width ratio would be perceived as most beautiful across all three groups, regardless of the degree of image processing. If supported, this would indicate that the observed preference is not an artifact of digital manipulation but rather reflects a general perceptual tendency.

\subsection{Experimental Design}
\label{Experimental design}

By manipulating the average faces within the conditions \textit{low}, \textit{medium}, and \textit{high}, a total of 21 distinct images were created. To present these faces, we generated all possible pairwise combinations of the seven variants within each condition. Thus, each face was compared to every other face in the same group, including the original, unaltered morphed average face. This yielded 42 unique pairwise comparisons per condition, amounting to a total of 126 comparisons across all three conditions. Each pair could be presented twice to control for order effects, resulting in $126 \cdot 2 = 252$ potential comparison screens.
Pairwise comparisons were chosen because, unlike in line-up designs, the manipulated width ratio remained imperceptible to participants. This minimized the risk of perceptual biases or social desirability effects, such as regression to the mean or an explicit focus on specific facial features that are socially associated with high beauty standards. 

However, comparisons between average faces from different conditions or identical face pairings were excluded. This decision was driven by methodological and ethical considerations that became apparent during the development of the experiment. Methodologically, each condition was based on a separate average face, and subtle differences in facial features, most notably skin tone, became apparent between these average faces. Direct comparisons across conditions would thus have introduced unwanted variation beyond the intended width ratio manipulation, compromising the internal validity of the study. Ethically, presenting participants with face pairs that differed in skin tone for beauty judgments risked introducing social-psychological confounds (e.g., response biases due to concerns about appearing racially biased) and could have been perceived as inviting racially charged evaluations. As this topic fell outside the study’s scope and risked causing participant discomfort, the design was restricted to within-condition comparisons (i.e., \textit{low}–\textit{low}, \textit{medium}–\textit{medium}, and \textit{high}–\textit{high} pairs). This ensured judgments were based solely on width ratio variations while upholding ethical research standards (these points are further discussed in Section 4.2).

\begin{figure}[!htbp]
\centering
\includegraphics[width=\textwidth]{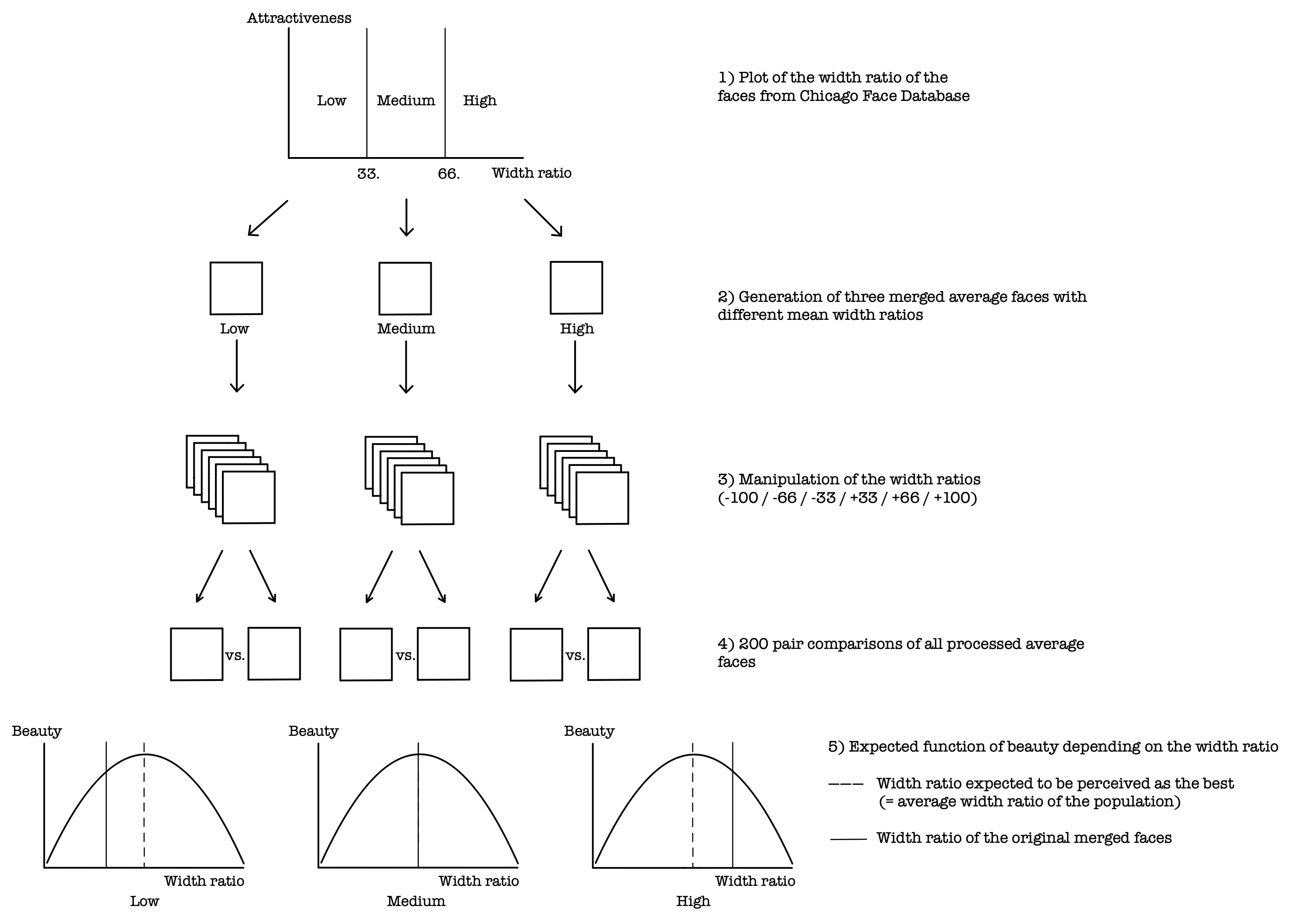}
\caption{Illustration of the methodological procedure: (1) plot of faces from the Chicago Face Database by width ratio, (2) generation of three averaged faces (\textit{low}, \textit{medium}, \textit{high} width ratios), (3) manipulation of each average face's width ratio using the Liquefy function in Adobe Photoshop, (4) pairwise comparisons of all manipulated faces within each condition, including two comparisons for each possible pair and one self-comparison per average face, (5) expected shift in perceived beauty towards the average face.}
\label{fig:Methodic procedure}
\end{figure}

To control for response tendencies, each condition included one trial in which the original (unaltered) average face was presented alongside itself. Participants were shown a total of 200 trials in a two-alternative forced-choice task paradigm. The remaining 71 trials (beyond the 126 structured pairings and 3 self-comparisons) consisted of randomly selected face pairings. These were included to increase the overall data volume while maintaining a reasonable study length to preserve participant motivation. Through this
procedure, all pairwise comparisons were expected to occur within approximately equal frequency across participants. Figure \ref{fig:Methodic procedure} provides an overview of the stimulus generation process and the composition of the experimental trials.

%Figure methodic procedure

\subsection{Procedure}
\label{Procedure}

The online experiment was made available via the University of Marburg's SONA system (Sona Systems, https://www.sona-systems.com) \nocite{SonaSystems}, where psychology students could access it through the university's study listing. In addition, the link to the experiment was distributed privately.

Before beginning the experiment, subjects were instructed to complete the study using a computer or tablet equipped with an external keyboard. They provided informed consent and read detailed instructions about the experimental procedure. Crucially, the instructions did not provide a precise definition of beauty. Given that the concept of beauty is subjective and can vary between individuals and cultures \citep{Barone2023}, participants were instead asked to respond as quickly and intuitively as possible. This approach was chosen to capture their personal, subjective understanding of beauty while minimizing deliberation based on explicit, normative socio-cultural standards \citep[cf.][]{Eagly1991} (the full instructions can be found in \ref{appendix: instructions} ).

After the experiment started, a fixation cross was displayed for 1000 ms. Pressing the space bar triggered the presentation of two images, randomly selected according to the procedure described above, shown on the left and right sides of the screen. Participants could select the left image by pressing the "F" key and the right image by pressing the "J" key. There was no time limit for making a decision. As soon as a response was given, the fixation cross reappeared, followed by the next comparison.

The experiment consisted of four blocks with 50 trials each. To support concentration, a break was provided after each block. This break lasted at least 15 seconds and could be terminated by pressing the space bar after this time had elapsed. Participants who were recruited via the university's SONA system could earn time-based credit points required for their academic progress in psychology studies. Otherwise, no compensation was provided.

\subsection{Statistical Analysis}
\label{ch:stats}
%Thurstone => ordnet aus den Vergleichen für jeden Stimulus einen reellen Wert zu

For statistical analysis, we used a Bayesian framework to model the pairwise
comparison data\footnote{\href{https://gitlab.uni-marburg.de/knoppbe/structural_beauty_analysis}{gitlab.uni-marburg.de/knoppbe/structural\_beauty\_analysis}}. Specifically, we employed the Bradley–Terry model, a slight
variation of the Thurstone model \citep{Cattelan2012}. This approach allowed us
to estimate a latent score $\mu_i$ for each stimulus image $i$, reflecting its
relative perceived beauty within each condition (\textit{low}, \textit{medium},
\textit{high}), based on the observed comparison data. For each comparison
between two stimuli $i$ and $j$, a probability $p_{ij} \in [0, 1]$ was computed
by applying a sigmoid function (Eq. \ref{eq:sigmoid} ) to the difference between
their latent scores $\mu_i - \mu_j$ (Eq. \ref{eq:p} ). Thus, $\mu_i > \mu_j$
means that stimulus $i$ is probably preferred to stimulus $j$. This probability
$p_{ij}$, together with the number $N_{ij}$ of comparisons between stimuli $i$
and $j$, parameterized a binomial distribution modeling the number of times
stimulus $i$ was preferred over stimulus $j$, denoted as $\text{win}_{ij}$ (Eq.
\ref{eq:obswin} ). Following \cite{Pallett2010}, we assumed a quadratic
relationship between the width ratio $wr_i$ of each stimulus and its
corresponding latent score $\mu_i$. We therefore estimated the parameters of a
parabola ($a$, $b$, $c$) and modeled latent scores as Gaussian-distributed,
with a mean given by this parabola and variance $\sigma$ (Eq.
\ref{eq:obsscore} ). Based on our hypothesis that average male faces are
perceived as most beautiful, the parabolas across the three conditions were
expected to converge toward a shared maximum near the average ratio. Together,
these components defined the posterior distribution from which we sought to
infer the width ratio perceived as most beautiful within each group. To
approximate this posterior, we employed the No-U-Turn
Sampler \citep{Hoffman2014} as implemented in \textit{PyMC5}
\citep{salvatier2016probabilistic}. We computed marginal likelihoods using
a bridge sampler \citep{gronau2017tutorial} implemented in \textit{PyMC5} \citep{salvatier2016probabilistic}.

\begin{align}
    \text{wr}^\star &\propto \text{Unif}(0.432, 0.5)\label{eq:wmax}\\
    c &\propto \text{Unif}(-3000, 0)\label{eq:c}\\
    b &= -2 c \cdot \text{wr}^\star\label{eq:b}\\
    a &= c \cdot \text{wr}^{\star 2}\label{eq:a}\\
    \sigma &\propto \mathcal{N}^+(1)\label{eq:sigma}\\
    \mu_i &\propto \mathcal N(a + b \cdot \text{wr}_i + c \cdot \text{wr}_i^2, \sigma)\label{eq:obsscore}\\
    \text{sigmoid}(x) &= \frac{1}{1 + \exp(-x)}\label{eq:sigmoid}\\
    p_{ij} &= \text{sigmoid}(\mu_i - \mu_j)\label{eq:p}\\
    \text{win}_{ij} &\propto \text{Binomial}(N_{ij}, p_{ij}\label{eq:obswin})
\end{align}

We chose our priors as follows: We set the maxima of the parabola to zero. 
The location $\text{wr}^\star$ of these maxima is drawn from a uniform distribution
between the minimal and maximal width ratios of all stimuli (Eq. \ref{eq:wmax} ).
The parabola parameters $a$ and $b$ are determined given $\text{wr}^\star$ and
$c$ (Eqs. \ref{eq:a} and \ref{eq:b} ). We uniformly draw $c$ from $-3000$ up to zero because we assume a convex
parabola (Eq. \ref{eq:c} ). The observation noise $\sigma$ is drawn from a half-normal distribution (Eq. \ref{eq:sigma} ).

%Results
\section{Results}
\label{Results}

We collected data from 97 participants comparing faces with differing width
ratios to check if a preference for averageness exists. Therefore, we developed a
Bayesian model enabling us to estimate the relationship between width ratio and
perceived beauty from 14,542 comparisons. Our data support a preference for
average width ratios.

\begin{figure}[!htbp]
\includegraphics[width=\textwidth]{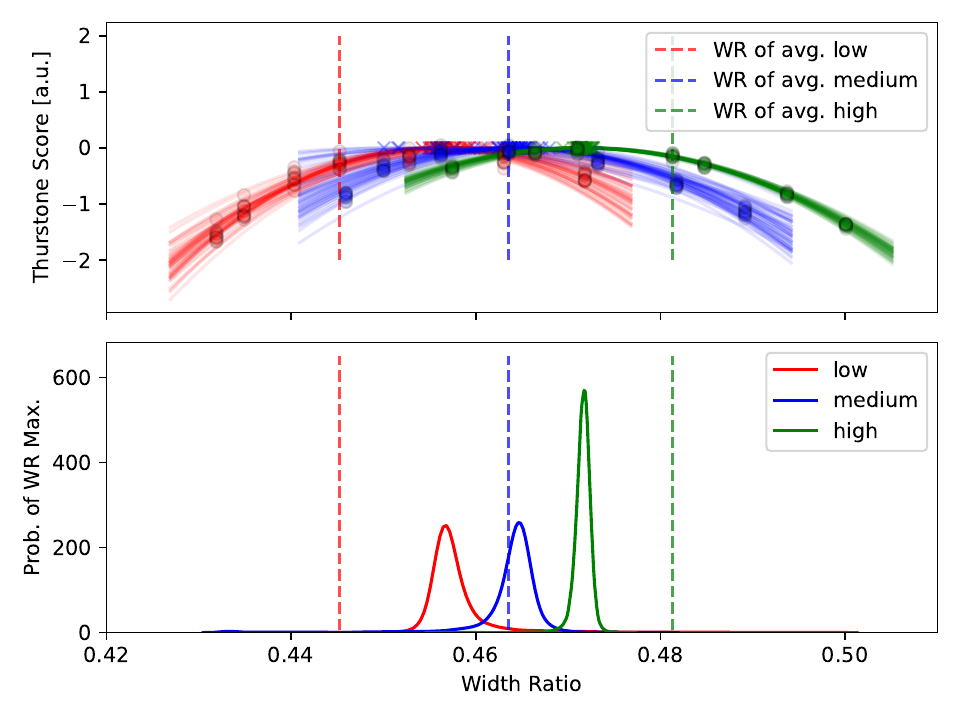}
    \caption{\textit{Top}: Circles show the estimated Thurstone scores for the seven stimuli for conditions \textit{low} (red), \textit{medium} (blue), and \textit{high} (green). Corresponding vertical dashed lines show the width ratio of the unmanipulated average image for each condition. Lines show 50 parabolas sampled from the posterior for each condition. Crosses show the maxima of the parabolas. \textit{Bottom}:  Probability of the beauty score maximum as a function of the width ratios.}
    \label{fig: result}
\end{figure}

Figure \ref{fig: result} displays the estimated relationship between width ratio and perceived beauty for each condition (\textit{low}, \textit{medium}, \textit{high}), based on the Bayesian model described in Section \ref{ch:stats}.
The upper part of the plot shows the estimated Thurstone score $\mu_i$ (Eq. \ref{eq:obsscore} ) for all stimulus images plotted against the corresponding width ratios.
The lines are parabolas corresponding to the parameters $\text{wr}^\star$ and $c$ (Eqs. \ref{eq:wmax} and \ref{eq:c} ) sampled from the posterior distribution.
The sampled maxima $\text{wr}^\star$ are depicted as $\times$, thus giving plausible guesses for optimal width ratio values, which are perceived as most beautiful based on our data\footnote{Note that our integrated Bayesian model enables us to interpolate
between width ratios used in stimulus images while ideally accounting for the uncertainty given the data and our model. This would not be the case if we had used a two-stage procedure, where the latent $\mu_i$ would be estimated before fitting the parabolas to these estimations.}.
The lower plot shows the kernel density estimate of the posterior maximum values. 
The vertical lines show the width ratios of the unaltered stimuli of each condition.

Note that because the pairwise comparisons were restricted to within-condition pairs (see Section \ref{Experimental design} ), the Thurstone scores are not comparable across conditions and can only be interpreted relative to the other stimuli within the same group. Descriptive trial counts for each pairwise comparison are provided in Tables \ref{table:low} to \ref{table:high} ( \ref{appendix:data} ).

Comparing the mean $\mu_i$ for each condition shows that only in the \textit{medium} condition does the unaltered image have the highest estimated Thurstone score. In the \textit{low} condition, increasing the width ratio by +33 yields the highest score, whereas in the \textit{high} condition, decreasing it by -33 yields optimal perceived beauty. This pattern is reflected in the estimated parabolas, whose maxima shifted toward the globally average width ratio, as shown in the density plot in Figure \ref{fig: result}.

%Figure: result

Next, we present the estimated optimal width ratios for each condition (\textit{low}, \textit{medium}, \textit{high}), and check whether it is more likely that the maxima are closer to the "medium" average width ratio compared to the "local" condition-specific average width ratio, which would indicate a preference for average width ratios. For this, we approximate the posterior probability density function (pdf) by the kernel density estimate of the maxima from MCMC samples, and evaluate the log pdf at the medium average $\text{wr}_\text{medium}$ and the local average $\text{wr}_\text{local}$. We perform a likelihood ratio comparison of these options by computing the log odds (see Figure \ref{fig:logodds} ):

\begin{align}
    \log \text{odds} &= \log \text{pdf}(wr_\text{medium}) - \log \text{pdf}(wr_\text{local}).
\end{align}

In the \textit{low} condition, the posterior exhibits a mean of $0.458 \pm 0.004$. 
    % The probability of observing the condition-specific mean under this posterior was very low ($p < 10^{-10}$), while the probability of observing the global mean (\textit{medium} reference) was marginally higher but still negligible ($p = 8.10 \times 10^{-12}$). 
The likelihood ratio comparison reveals that the global average is $95$ times more probable as the inferred maximum than the condition-specific average, indicating a shift towards the global average.
The posterior mean of the \textit{medium} condition is $0.464 \pm 0.004$.
For the \textit{high} condition, the posterior mean is $0.472 \pm 0.001$. 
    %The probability of observing the condition-specific mean was vanishingly small ($p = 2.48 \times 10^{-25}$), and also the probability for the global mean remained low ($p = 8.69 \times 10^{-5}$). 
The likelihood ratio indicates that the global average is more than $10^{230}$ times more probable as the inferred maximum than the condition-specific average.
	% The \textit{medium} condition, which served as the global reference, showed close alignment between the posterior and the observed data, with $\mu = 0.464$ ($\sigma = 0.001$). Here, the probability of observing the global mean was 422.6, reflecting a near-perfect fit. This suggests that the posterior distribution accurately captures the empirical optimum in this condition.
These results reveal that within each condition, the posterior supports a preference for the global average, while in the \textit{medium} condition the global average is similar to the local average.

% We evaluated the model by sampling from its posterior predictive distribution. For the \textit{low}, \textit{medium}, and \textit{high} conditions, respectively, $43\%$, $53\%$ and $47\%$ of the observed pairwise wins $\text{win}_{ij}$ fell between the 25th and 75th percentiles of the predicted samples. These proportions suggest that the model provides a reasonable fit to the data.

\begin{figure}[!htbp]
\includegraphics[width=\textwidth]{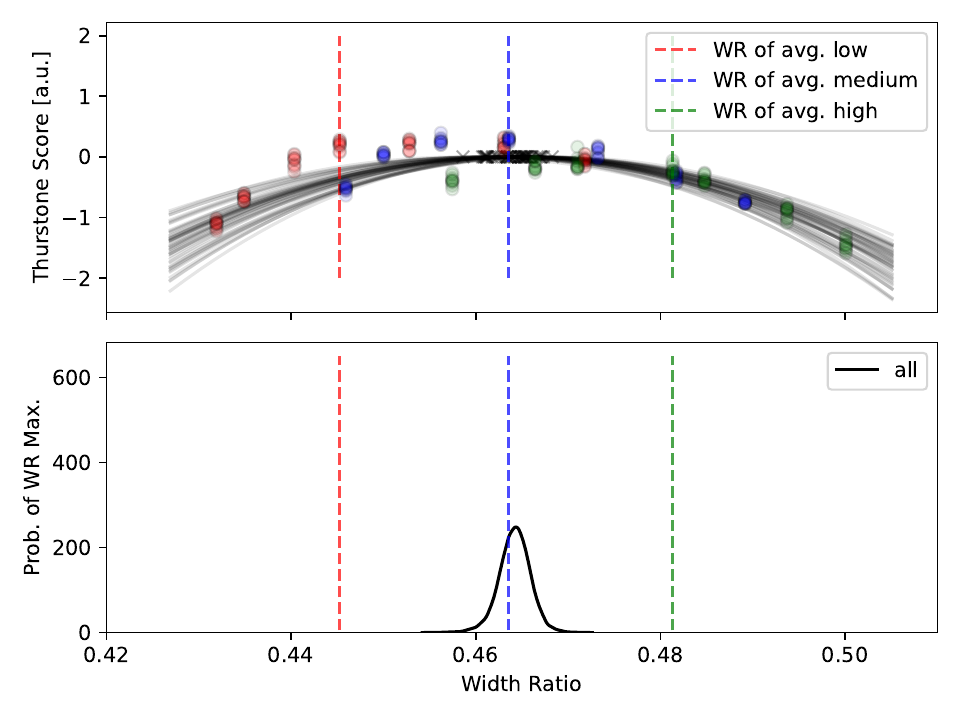}
    \caption{H0 model with parabola shared across conditions. See explanation in Figure \ref{fig: result}. The only difference is that black lines show 50 parabolas (shared across conditions) sampled from the posterior. }
    \label{fig:h0}
\end{figure}

We evaluated whether a model with a shared parabola for all conditions would be sufficient to explain our data. Therefore, we compared the marginal likelihood of the model presented above with distinct parabolas for the different conditions (H1) and a model with a parabola shared across conditions (H0, see Figure \ref{fig:h0} ). The results are shown in Table \ref{tab:comparison}: They show a clear preference for the model with condition-specific parabolas.

\begin{table}[h]
\centering
\begin{tabular}{lc}
\toprule
    Model & Marg. LLH [nats] \\
\midrule
H0 & -254.7 \\
H1 & -247.4 \\
\bottomrule
\end{tabular}
\caption{Result of Bayesian model comparison.}
\label{tab:comparison}
\end{table}

%Discussion
\section{Discussion}
\label{Discussion}

%Zusammenfassung

This experiment systematically manipulated the facial width ratio, defined as the interocular distance relative to the total width of the face, and assessed its effect on perceived beauty using pairwise comparisons. Participants indicated which of two simultaneously presented faces they found more beautiful, providing a measure of the relative perception of subjective beauty. The results indicate that the global mean width ratio better explains the perceptual data across conditions than condition-specific means, supporting the hypothesis that faces closer to the average width ratio are perceived as most beautiful.

\subsection{Methodological Implications of Image Manipulation}

To manipulate the width ratios, multiple average faces with different baseline proportions were created by morphing together 96 or 97 male faces each from the Chicago database \citep{Ma2015}. This averaging technique made it possible to generate realistic composite faces that could be manipulated in a way such that they differed only in their width ratios. Holding these additional features constant minimized noise factors and isolated the effect of the manipulated ratio.

Even though inter-conditional comparisons were excluded from the experimental
design, the use of average faces with different width ratios allowed the
results from the \textit{low} and \textit{high} conditions to serve as a control for the
main effect expected in the \textit{medium} condition. Accordingly, it was
hypothesized that the parabolas representing the relationship between beauty
and width ratios would converge at a shared maximum near the average ratio.
Across all three conditions, the maxima
consistently shifted toward the overall average width ratio, indicating a
robust tendency toward averageness in perceived facial beauty. This constitutes
a particular strength of the study, as it enhances the internal validity of the
experiment and provides transparency regarding the processed nature of the
stimuli.

%1) Bildbearbeitung
However, the point of maximum beauty did not lie exactly at the mean ratio, and a residual difference between the beauty maxima across the three initial conditions remained. This incomplete convergence in the \textit{low} and \textit{high} conditions may plausibly be attributed to the degree of image manipulation in the conditions that revealed greater deviations from the overall mean. We confirmed this with Bayesian model comparison, where we showed that a model with condition-specific parabolas explains the data better than a model with a single parabola.
%2) eingeschränkte Generalisierbarkeit, weil Gesichter existieren nicht wirklich
Still, it must be acknowledged that none of the morphed faces presented in this experiment represents an actual person, limiting the generalizability of these findings to real, unaltered faces.

\subsection{Methodological Adjustments Due to Ethical Concerns}
\label{discussion: methodological_adjustments}
%4) die Gruppierung hatte allerdings Nachteile: keine Quervergleiche mgl., weil aus ethischen Gründen Quervergleiche von Gesichtern mit verschiedenen Hautfarben ausgeschlossen wurden -> drei Bedingungen sollten aber erhalten bleiben => deshalb Diskussion, welche Effekte die Unterschiede zwischen den Bedingungen in Bezug auf die wahrgenommene Schönheit haben könnten

As detailed in Section \ref{Experimental design}, the decision to restrict comparisons to within-condition pairs (\textit{low}–\textit{low}, \textit{medium}–\textit{medium}, \textit{high}–\textit{high}) was a direct consequence of subtle skin-tone differences between the average faces of the three conditions. This necessary design choice limits direct comparability of beauty scores across conditions.

Methodologically, using three separate average faces helped control for potential Photoshop-related artifacts and ensured that the preference for more average width ratios was not an artifact of image processing. Ethical considerations also played a role: presenting participants with face pairs differing in skin tone could have implicitly encouraged racially biased evaluations, which was not the focus of this study.

Consequently, a single, universal optimal width ratio for male faces cannot be determined. However, converging trends suggest that an optimal ratio likely lies within the range of $0.457$ and $0.464$, close to the $46\%$ reported for female faces by \cite{Pallett2010}. Future studies could generate a single average face from all male faces to allow cross-condition and cross-sex comparisons, removing between-condition variability in features like skin tone and allowing for unambiguous comparisons of male face average ratios with the values reported for female faces \citep{Pallett2010}.

    %Nonetheless, the original decision to employ three distinct average faces served a critical methodological purpose, as it controlled for potential confounds introduced by Photoshop-based manipulations within any one image set. It ensured that the observed preference for more average ratios was replicable across different facial contexts and was not an artifact of average image processing. As such, this step remains essential for validating the manipulation itself and should be considered a valuable methodological check in studies investigating the perception of manipulated facial features.

\subsection{Naturalness and Ethnicity as Possible Confounding Variables}

%5) natürlichere Gesichter durch Blurring
Morphing multiple individual faces into average composites introduces a slight blurring effect, which enhances natural-looking manipulations of width ratios and minimizes the salience of individual features. This may have reduced perceived naturalness, potentially explaining why the most beautiful width ratio was not perfectly consistent across all conditions.

%6) Co-Variation durch Ethnizität
Another potential confound is ethnicity. Slight variations in skin tone between composite images may reflect sampling biases within the database or structural differences associated with ethnic background \citep[cf.][]{Naik2022, Jothishankar2019}. Overrepresentation of certain ethnic groups in subsets of faces with particularly low or high width ratios could unintentionally influence beauty ratings based not purely on geometric proportions but on associated group characteristics. %This might also explain the slight deviations observed in Figure \ref{fig: result}: Although a general trend toward the mean width ratio was evident, the points of maximum perceived beauty across the three conditions did not coincide in the exact same width ratios. To verify or rule out such a hypothesis,
Future studies should consider ethnicity as an additional variable, with care for cultural sensitivity and ethical implications \citep{Dimitrov2023, Frevert2014}.

\subsection{Socio-Cultural and Ethical Implications}

Given the importance of gender-dimorphic characteristics in beauty perception, the investigation of male and female genders appears self-evident. However, the binary gender assumption of this study cannot fully capture human gender diversity. Non-binary individuals were not assessed, but this remains an important topic for future research. 

Beauty has societal relevance, being associated with social status, professional success, and income \citep[e.g.,][]{Hamermesh2013, Hosoda2003, Mobius2006}. Research in this area must remain critically aware of privileges and forms of discrimination related to beauty as a social construct. Variable factors such as clothing style and posture, can influence perceived attractiveness \citep{Tzschaschel2019, Hester2023}, and cultural imprinting, social norms, and role models play important roles in shaping beauty perception \citep[e.g.,][]{Blais2021, Frevert2014, Dimitrov2023}. These multifaceted influences underscore the importance of nuanced and ethically sensitive research in this field \citep{Zebrowitz2008}.

\subsection{Width and Length Ratios}

This study does not address whether the observed preference for average proportions also applies to other facial ratio measures, such as the eye-to-mouth distance relative to total face length \citep{Pallett2010}, or interactions between width and length ratios. Investigating such effects lies beyond the scope of the present study but remains a question for future research.

\section{Conclusion}
%Zusammenfassung, Einordnung in größeren Kontext, Nutzen

This study provides empirical support for the hypothesis that average facial proportions, specifically the ratio between interocular distance and overall facial width, are perceived as more attractive. These findings align with and extend existing theoretical frameworks: From a cognitive perspective, average faces are processed more fluently due to their resemblance to internalized prototypes \citep{Langlois1990, Pallett2010}; from an evolutionary perspective, averageness may signal genetic fitness and health, providing an adaptive advantage to preferring such faces \citep{Pallett2010, Rhodes2006}.

Beyond replicating the preference for averageness, the study highlights the methodological importance of controlling for image-processing artifacts and second-level factors such as ethnicity. The observed shifts in beauty maxima suggest that representations of various ethnic-specific facial prototypes may moderate perceptual fluency and preference \citep{Alter2009, Langlois1990, Arkes1991, Juravle2024}. Consequently, future research should move beyond WEIRD (Western, educated, industrialized, rich, democratic) populations \citep{Henrich2010} and consider ethnicity and cultural diversity as moderating variables, while maintaining methodological rigor and ethical sensitivity \citep{Zebrowitz2008, Frevert2014, Dimitrov2023}.

In sum, this study strengthens the evidence for a geometric basis of facial beauty, while highlighting the complex interplay of cognitive, evolutionary, and socio-cultural factors in shaping aesthetic preferences \citep{Raggio2022}.

\section{Applications and Outlook}
\label{Applications}

Although the precise mechanisms underlying the perception of beauty remain to be fully understood, the consistent preference for averageness observed in this study suggests several practical and theoretical applications.

In aesthetic medicine and cosmetics, knowledge of average facial proportions can inform culturally sensitive interventions and product development \citep{Borelli2010, Adamson2006, Dimitrov2023, Laughter2023, Hashim2017}.
In the context of artificial intelligence and virtual agent design, integrating geometric averages into the design of artificial faces could enhance user acceptance and emotional resonance in human-computer interaction \citep[e.g.,][]{Jackson1995}.

To deepen our theoretical understanding, future research should employ longitudinal and cross-cultural designs to examine how variables such as age, social context, and sociocultural change (e.g., globalization, migration) modulate internalized facial prototypes \citep{Sarwer2003, Jones2011, Little2011}. Crucially, such studies must prioritize methodological transparency (e.g., controlling for image-editing artifacts) and ethical awareness to avoid reinforcing social biases when studying a construct as socially consequential as beauty \citep{Zebrowitz2008, Frevert2014, Eagly1991}.

Taken together, this study not only confirms a geometric foundation for facial beauty but also outlines a pathway for ethically grounded and culturally inclusive approach to research and clinical practice, emphasizing the need for robust methodological frameworks to systematically examine the perception of beauty as a culturally embedded psychological construct.

\section*{Data Availability}

We provide anonymized raw data, processing and analysis code here: \href{https://gitlab.uni-marburg.de/knoppbe/structural_beauty_analysis}{gitlab.uni-marburg.de/knoppbe/structural\_beauty\_analysis}

\section*{Acknowledgements}

This work was funded by DFG, IRTG1901 - The brain in action, and SFB-TRR 135 - Cardinal mechanisms of perception. The authors thank 
Bente Bues, Johanna Hänel, Paula Härterich, Julia Hinske, Franka Nimtz, Hannah Rudolph, Ece Sahin, Justin Triller, Jessica Witzel and
Ruth Zimmermann for their contributions to the design study and data collection.

\section*{Conflict of Interest}

The authors declare no conflict of interest.

\section*{Author Contributions}

TT conceived and designed the study, collected the data, and drafted the manuscript. BK conceived and designed the study and analyzed and interpreted the data. DE analyzed and interpreted the data. All authors revised the work and approved the final version for publication.

\printendnotes

% Submissions are not required to reflect the precise reference formatting of the journal (use of italics, bold etc.), however it is important that all key elements of each reference are included.
\bibliography{structural_beauty}

\hspace{1em}

\newpage

%Appendix
\appendix

\section{Data Tables}
\label{appendix:data}

\begin{table}[!htb]
\centering
\begin{tabular}{llrr}
\toprule
 &  & win1 & win2 \\
1 & 2 &  &  \\
\midrule
\multirow[t]{6}{*}{-100} & 33 & 47 & 167 \\
 & 100 & 61 & 162 \\
 & 66 & 55 & 188 \\
 & -33 & 50 & 170 \\
 & 0 & 47 & 183 \\
 & -66 & 80 & 134 \\
\cline{1-4}
\multirow[t]{5}{*}{33} & 100 & 133 & 106 \\
 & 66 & 110 & 110 \\
 & -33 & 147 & 87 \\
 & 0 & 117 & 123 \\
 & -66 & 174 & 64 \\
\cline{1-4}
\multirow[t]{4}{*}{100} & 66 & 99 & 130 \\
 & -33 & 118 & 128 \\
 & 0 & 108 & 116 \\
 & -66 & 153 & 82 \\
\cline{1-4}
\multirow[t]{3}{*}{66} & -33 & 128 & 103 \\
 & 0 & 136 & 95 \\
 & -66 & 178 & 72 \\
\cline{1-4}
\multirow[t]{2}{*}{-33} & 0 & 92 & 141 \\
 & -66 & 155 & 79 \\
\cline{1-4}
0 & -66 & 176 & 60 \\
\cline{1-4}
\bottomrule
\end{tabular}

    \caption{Comparison data for condition \textit{low}: Multi-index with labels 1 \& 2 show the compared stimuli labeled by the manipulated width ratio in percent. \textit{win1} shows the number of trials in which stimulus 1 was preferred over stimulus 2, and \textit{win2} vice versa.}
\label{table:low}
\end{table}

\newpage

\begin{table}[!htb]
\centering
\begin{tabular}{llrr}
\toprule
 &  & win1 & win2 \\
1 & 2 &  &  \\
\midrule
\multirow[t]{6}{*}{-33} & -100 & 160 & 75 \\
 & 100 & 164 & 61 \\
 & 33 & 135 & 95 \\
 & 66 & 165 & 86 \\
 & -66 & 125 & 93 \\
 & 0 & 118 & 108 \\
\cline{1-4}
\multirow[t]{5}{*}{-100} & 100 & 138 & 99 \\
 & 33 & 73 & 153 \\
 & 66 & 103 & 124 \\
 & -66 & 69 & 181 \\
 & 0 & 76 & 157 \\
\cline{1-4}
\multirow[t]{4}{*}{100} & 33 & 75 & 163 \\
 & 66 & 84 & 142 \\
 & -66 & 64 & 171 \\
 & 0 & 55 & 163 \\
\cline{1-4}
\multirow[t]{3}{*}{33} & 66 & 129 & 92 \\
 & -66 & 117 & 102 \\
 & 0 & 114 & 121 \\
\cline{1-4}
\multirow[t]{2}{*}{66} & -66 & 85 & 147 \\
 & 0 & 79 & 150 \\
\cline{1-4}
-66 & 0 & 77 & 142 \\
\cline{1-4}
\bottomrule
\end{tabular}

    \caption{Comparison data for condition \textit{medium}: See Table \ref{table:low} for description.}
\label{table:medium}
\end{table}

\newpage

\begin{table}[!htb]
\centering
\begin{tabular}{llrr}
\toprule
 &  & win1 & win2 \\
1 & 2 &  &  \\
\midrule
\multirow[t]{6}{*}{33} & 66 & 155 & 88 \\
 & -33 & 108 & 145 \\
 & -66 & 103 & 132 \\
 & -100 & 121 & 103 \\
 & 100 & 159 & 66 \\
 & 0 & 114 & 131 \\
\cline{1-4}
\multirow[t]{5}{*}{66} & -33 & 70 & 166 \\
 & -66 & 68 & 168 \\
 & -100 & 82 & 150 \\
 & 100 & 149 & 75 \\
 & 0 & 82 & 145 \\
\cline{1-4}
\multirow[t]{4}{*}{-33} & -66 & 121 & 100 \\
 & -100 & 134 & 106 \\
 & 100 & 183 & 52 \\
 & 0 & 118 & 106 \\
\cline{1-4}
\multirow[t]{3}{*}{-66} & -100 & 121 & 105 \\
 & 100 & 178 & 56 \\
 & 0 & 125 & 91 \\
\cline{1-4}
\multirow[t]{2}{*}{-100} & 100 & 167 & 58 \\
 & 0 & 103 & 131 \\
\cline{1-4}
100 & 0 & 40 & 173 \\
\cline{1-4}
\bottomrule
\end{tabular}

    \caption{Comparison data for condition \textit{high}: See Table \ref{table:low} for description.}
\label{table:high}
\end{table}

\newpage

\begin{figure}[!htbp]
\includegraphics[width=\textwidth]{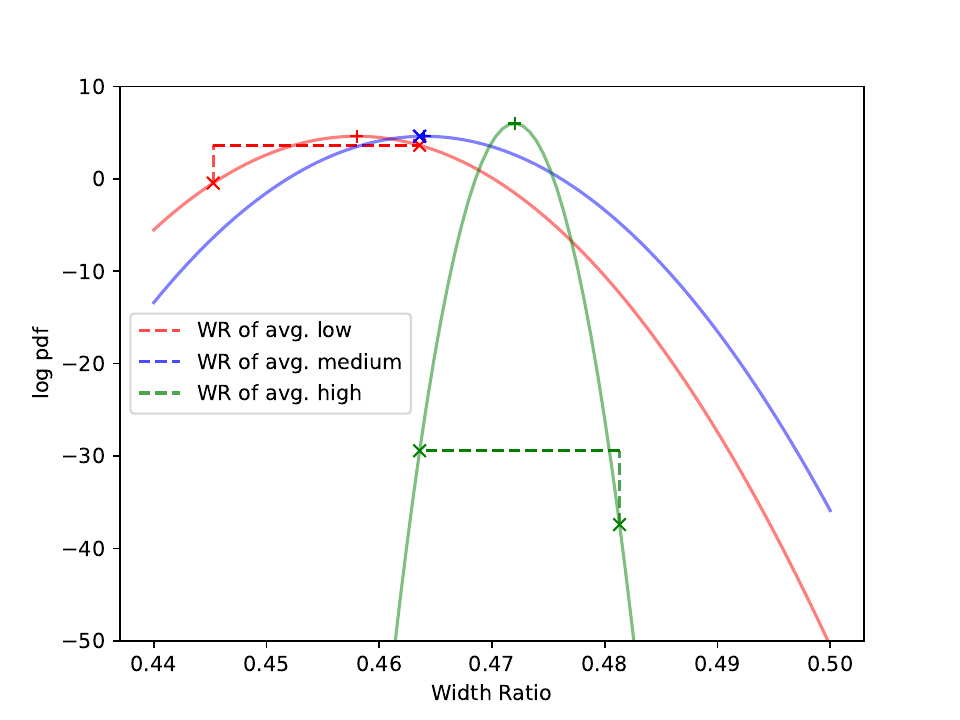}
    \caption{Illustration of likelihood ratio comparison: Lines show the kernel
    density estimation approximation of the posterior pdf of inferred optimal width values for each condition.
    The $+$ markers show the modes of the distributions, while $\times$ show the logpdf of the "local" average 
    stimulus. The vertical line shows the log odds: The difference between the logpdf value at
    the medium stimulus width ratio and the logpdf value at the "local" average width ratio (connected
    by the horizontal line).
    }
    \label{fig:logodds}
\end{figure}

%Additinoal information
\section{Additional Information on the Experiment}
\label{Additional information on the experiment}

\subsection{Declaration of Consent}
\label{appendix: declaration of consent}

In dieser Studie untersuchen wir die Wahrnehmung von Schönheit. Die Studie ist Teil des Experimentalpraktikums "Psychophysikalische Evaluation von Deep Learning Modellen" (2022) am Fachbereich für Psychologie und wird zu Lernzwecken durchgeführt.
Mit meiner Einwilligung erkläre ich, dass ich die Information zur Studie gelesen und verstanden habe. Ich habe keine weiteren Fragen bezüglich der Studie und der verbundenen Risiken. Ich weiß, dass eine Ablehnung oder ein Rücktritt von der Studie keine negativen Konsequenzen für mich hat. Ich entscheide mich freiwillig für die Teilnahme an der Studie und willige in die Verarbeitung und ggf. Veröffentlichung meiner pseudonomisierten Daten ein. Ich weiß, dass die von mir erhobenen Daten aufgrund der Pseudonomisierung nicht mehr gelöscht werden können und stimme diesem zu.
Wenn Sie Fragen zur Forschung oder zu den Rechten der Studienteilnehmer*innen hast, zögern Sie bitte nicht, die Forscher über diese E-Mail zu kontaktieren: knoppbe@staff.uni-marburg.de
Um einzuwilligen, klicken Sie bitte "Ich stimme zu."
Wenn Sie nicht einwilligen, können Sie einfach das Browserfenster schließen, um das Experiment zu beenden.

\subsection{Instructions}
\label{appendix: instructions}

Herzlich Willkommen zu unser Studie!
Im Folgenden präsentieren wir Ihnen jeweils zwei Bilder mit jeweils einem Gesicht, bei denen Sie entscheiden sollen, welches der beiden Sie als schöner empfinden. Schönheit ist dabei im Sinne von Ästhetik zu verstehen. Wenn Sie das linke Gesicht für schöner halten, drücken Sie bitte die Taste "F", bei dem rechten Gesicht bitte die Taste "J". Es gibt keine richtigen und falschen Antworten. Schönheit liegt im Auge des Betrachters. Wir möchten Sie bitten, so schnell und intuitiv wie möglich zu antworten.
Bitte bearbeiten Sie die Studie allein an einem ruhigen und ungestörten Ort, wenn möglich an einem Computer. Eine Teilnahme mit einem Tablet ist aus technischen Gründen nur mit externer Tastatur möglich.
Die Unschärfe der Bilder ist beabsichtigt und nicht auf etwaige technische Probleme bzw. Sehkraft zurückzuführen.
Die Untersuchung wird ca. 15 Minuten Ihrer Zeit beanspruchen. Es gibt insgesamt 4 Blöcke mit jeweils 50 Durchgängen. Im Anschluss an jeden Block folgt eine Pause. Mit einem Klick auf die Leertaste können Sie die Pause eigenständig beenden und es beginnt automatisch der nächste Block.
Sofern Sie Student*in der Philipps-Universität Marburg sind, können Sie sich am Ende der Studie 0,25 Versuchspersonenstunden über SONA anrechnen lassen.

\subsection{Debriefing After the Experiment}
\label{appendix: debriefing}

Falls Sie sich während des Experiments gefragt haben, wie sich die Gesichter jeweils unterscheiden, möchten wir Sie zuletzt auch darüber aufklären. Mit dieser Studie wollen wir herausfinden, inwiefern das Verhältnis des Augenabstands zur Gesichtsbreite die Wahrnehmung der Schönheit männlicher Gesichter beeinflusst. Dafür haben wir männliche Gesichter verschiedenster Herkunft zu Durchschnittsgesichtern zusammengefasst, wodurch auch die Unschärfe in den dargestellten Bildern entstanden ist. Anschließend haben wir jeweils den Augenabstand manipuliert.
Wir bedanken uns erneut recht herzlich für Ihre Teilnahme!

\section{Statement on the Use of Artificial Intelligence}
\label{appendix: ai}

In addition to the AI-supported applications described in the Methods section, Chat-GPT (GPT-4 turbo; OpenAI, 2024) \nocite{OpenAI2024} and Meta Llama (3.1 8G; Meta, 2023) were used for grammar correction and to improve text readability. All AI-generated suggestions were manually verified for content accuracy, and the authors assume full responsibility for the final text.

\end{document}